\documentclass[twocolumn,showpacs,aps,prl,superscriptaddress]{revtex4}

\usepackage{graphicx}
\usepackage{dcolumn}
\usepackage{amsmath}
\usepackage{epsfig}

\input pubboard/babarsym

\newcommand{\BABARPubYear}    {05}
\newcommand{\BABARPubNumber}  {18}

\newcommand{\SLACPubNumber} {11278}
\newcommand{\LANLNumber} {0506036}

\newcommand{\Denu}{\ensuremath{D\xspace\electron\nub}}
\newcommand{\shmax}{\ensuremath{s_{\mathrm{h}}^{\mathrm{max}}}}
\newcommand{\shmaxz}{\ensuremath{\tilde{s}_{\mathrm{h}}^{\mathrm{max}}}}

\newcommand{\btouenu}{\ensuremath{\Bbar\to X_u\electron\nub}}
\newcommand{\btocenu}{\ensuremath{\Bbar\to X_c\electron\nub}}
\newcommand{\btoclnu}{\ensuremath{\Bbar\to X_c\ell\overline{\nu}_{\ell}}}

\newcommand{\GEVCC}{\ensuremath{{\mathrm{\,Ge\kern -0.1em V^2\!/}c^4}}\xspace}
\newcommand{\GEV}{\ensuremath{{\mathrm{\,Ge\kern -0.1em V^2}}}\xspace}
\newcommand{\etal}{{\em et al.}}

\newcommand{\Bunfa}{\ensuremath{\Delta\mathcal{B}(1.9,3.5)=(5.29\pm 0.44\pm 0.72)\times 10^{-4}}}
\newcommand{\Bunfb}{\ensuremath{\Delta\mathcal{B}(2.0,3.5)=(4.41\pm 0.42\pm 0.42)\times 10^{-4}}}
\newcommand{\Bunfc}{\ensuremath{\Delta\mathcal{B}(2.1,3.5)=(3.68\pm 0.43\pm 0.36)\times 10^{-4}}}

\newcommand{\Vubb}{\ensuremath{|V_{ub}|=(4.41\pm 0.30\ {}^{+0.65}_{-0.47}\pm 0.28)\times 10^{-3}}}

\begin{document}


\preprint{\babar-PUB-\BABARPubYear/\BABARPubNumber} 
\preprint{SLAC-PUB-\SLACPubNumber} 

\begin{flushleft}
\babar-PUB-\BABARPubYear/\BABARPubNumber\\
SLAC-PUB-\SLACPubNumber\\
hep-ex/\LANLNumber\\
\end{flushleft}

\begin{flushright}
\end{flushright}

\title {\large\bf Determination of \boldmath$\Vub$ from Measurements of the Electron and 
Neutrino Momenta in Inclusive Semileptonic \B\ Decays}

%
\author{B.~Aubert}
\author{R.~Barate}
\author{D.~Boutigny}
\author{F.~Couderc}
\author{Y.~Karyotakis}
\author{J.~P.~Lees}
\author{V.~Poireau}
\author{V.~Tisserand}
\author{A.~Zghiche}
\affiliation{Laboratoire de Physique des Particules, F-74941 Annecy-le-Vieux, France }
\author{E.~Grauges}
\affiliation{IFAE, Universitat Autonoma de Barcelona, E-08193 Bellaterra, Barcelona, Spain }
% ...
\collaboration{The \babar\ Collaboration}
\noaffiliation

\date{\today}

\begin{abstract}
We present a determination of the CKM matrix element \Vub\ based on
the analysis of semileptonic \B\ decays from a sample of $88$ million
\FourS\ decays collected with the \babar\ detector at the \pep2\
\epem\ storage ring.  Charmless semileptonic \B\ decays are selected
using measurements of the electron energy and the invariant mass squared
of the electron-neutrino pair.  
We obtain $\Vubb$, where the errors represent
experimental uncertainties, heavy quark parameter uncertainties,
and theoretical uncertainties, respectively.

Note: this document includes corrections to the original
publication of this work~\cite{vubprl}.  These corrections, which
affect the calculated efficiencies and quantities derived from
them, will appear in an erratum.
\end{abstract}

\pacs{13.20.He, 12.15.Hh, 14.40.Nd} 
\maketitle  

The study of the weak interactions of quarks has played a crucial role
in the development of the Standard Model (SM), which embodies our
understanding of the fundamental interactions.  The increasingly
precise measurements of $CP$ asymmetries in \B\ decays allow stringent
experimental tests of the SM mechanism for $CP$ violation
via the complex phase in the Cabibbo-Kobayashi-Maskawa (CKM) matrix~\cite{ref:KM}.
Improved determinations of \Vub, the coupling strength of the \b\ quark 
to the \u\ quark, will improve the sensitivity of these tests.

Two observables have been used to determine \Vub\ from inclusive 
semileptonic \B\ decays: the endpoint of the lepton momentum
spectrum~\cite{ref:endp-meas}
and the mass of the accompanying hadronic system~\cite{ref:mx-meas}.
In this paper,
semileptonic $\btouenu$ decays are selected using 
a novel approach based on
simultaneous requirements for the electron energy, $E_{e}$, and 
the invariant mass squared of the $\electron\nub$ pair, 
$q^2$~\cite{ref:shmax}.  The neutrino 
4-momentum is reconstructed from the visible 4-momentum and knowledge
of the \epem\ initial state.  The dominant charm background is
suppressed by selecting a region of the $q^2$-$E_{e}$ phase space
where correctly reconstructed $\btocenu$ events are kinematically
excluded.  Background contamination in the signal region is due to
resolution effects and is evaluated in Monte Carlo (MC) simulations.  
Theoretical calculations are applied to the measured 
$\btouenu$ partial rate to determine \Vub, 
the precision of which is limited mostly by our current knowledge 
of the \b-quark mass, $m_b$.

The data used in this analysis were collected with the 
\babar\ detector~\cite{ref:babar} at the \pep2\ asymmetric-energy 
\epem\ storage ring.
The data set consists of $88.4$ million $\BB$ pairs
collected at the \FourS\ resonance,
corresponding to an integrated luminosity of $81.4\invfb$ at
$\sqrt{s}=10.58 \gev$.  An additional $9.6\invfb$ of data were collected at 
center-of-mass energies $20\mev$ below the \BB\ threshold.  
Off-resonance data are used to subtract the
non-\BB\ contributions from the data collected at the \FourS\ resonance. 
To do so, the off-resonance data are scaled according to the
integrated luminosity and the energy dependence
of the QED cross-section, and the particles are boosted to 
the \FourS\ resonance energy.
Throughout this paper, all kinematic variables are given in 
the \FourS\ rest frame unless stated otherwise.

The simulation of charmless semileptonic \B\ decays used in optimizing
the analysis and determining reconstruction efficiencies is based on
the Heavy Quark Expansion (HQE) including
$\mathcal{O}(\alpha_S)$ corrections~\cite{ref:NdF}.  
This calculation produces a continuous spectrum of hadronic masses, $m_X$. 
Subsequent hadronization is simulated using JETSET down to 
$2 m_\pi$~\cite{ref:SJostrand}.  Decays to low-mass hadrons ($\pi$,
$\eta$, $\rho$, $\omega$, $\eta^{\prime}$) are simulated separately
using the ISGW2 model~\cite{ref:ISGW2}, and mixed with the
non-resonant states so that the $m_X$, $q^2$ and $E_e$ spectral
distributions correspond as closely as possible to the HQE
calculation.

Hadronic events containing an identified electron with energy
$2.1\gev<E_e<2.8\gev$ are selected.  Radiative Bhabha events rejected 
using the criteria given in Ref.~\cite{ref:elecID} and electrons 
from $\jpsi\to\epem$ decays are vetoed.  The total
visible 4-momentum, $p_\mathrm{vis}$, is determined using charged tracks
emanating from the collision point, identified pairs of charged tracks
from $\KS\to\pip\pim$, $\Lambda\to\proton\pim$ and $\gamma\to\epem$,
and energy deposits in the electromagnetic calorimeter.  Each charged
particle is assigned a mass hypothesis based on particle
identification information.  Calorimeter clusters unassociated with a
charged track and with a lateral energy spread consistent with
electromagnetic showers are treated as photons.  

Additional requirements are made to improve the quality of the neutrino 
reconstruction and suppress contributions from $\epem\to\qqbar$ 
continuum events.  
We form the
missing 4-momentum, $p_{\mathrm{miss}}=p_{\epem}-p_{\mathrm{vis}}$,
where $p_{\epem}$ is the 4-momentum of the initial state.
For each event we require
(1) no additional identified \electron\ or \mmu; (2)
$-0.95<\cos\theta_{\mathrm{miss}}<0.8$, where
$\theta_{\mathrm{miss}}$ is the polar angle of the missing
3-momentum; (3)
$0.0\gev<E_{\mathrm{miss}}-|{\bf p}_{\mathrm{miss}}|<0.8\gev$, where
$E_{\mathrm{miss}}$ is the missing energy in the event; (4)
$|{\bf p}_\mathrm{miss}|<2.5\gev$ and (5)
$|\cos\theta_T|<0.75$, where $\theta_T$ is the angle between
the electron momentum and the thrust vector
of the remaining particles in the event. 

The measured $|{\bf p}_{\mathrm{miss}}|$ differs from the true neutrino
momentum due to additional particles that escape detection.  Therefore,
a bias correction,
${\bf p}_\nu={\bf p}_{\mathrm{miss}}(0.804 - 0.078/|{\bf p}_{\mathrm{miss}}|)$,
is derived from the simulation.  Since the resolution on
$|{\bf p}_{\mathrm{miss}}|$ is superior to that of 
$E_{\mathrm{miss}}$,
we set $p_\nu = ({\bf p}_\nu, |{\bf p}_\nu|)$ and $q^2=(p_e+p_{\nu})^2$.
Defining $\eta_{\pm}=\sqrt{(1 \pm \beta)/(1 \mp\beta)}$,
where $\beta\simeq 0.06$ is the velocity of the \B\ meson in the \FourS\ frame,
the maximum kinematically allowed hadronic mass squared
for a given $E_e$ and $q^2$ is
$\shmax = m_{B}^2 +q^2 - 2m_{B}( E_{e}\eta_- + q^2\eta_+/{4 E_{e}})$
for $\pm 2E_{e} > \pm \sqrt{q^2}\,\eta_\pm$,
and $\shmax= m_{B}^2 + q^{2} -2 m_{B}\sqrt{q^2}$
otherwise.  We require $\shmax< 3.5\gev^2 \approx m_{D^0}^2$;
no \btocenu\ decays can have values of
$\shmax$ below this limit before accounting for
resolution.  
The requirements on $E_e$ and $\shmax$ and
criteria (1)--(5) were chosen to minimize the total
(experimental and theoretical) 
expected uncertainty $\sigma(\Vub)/\Vub$.

\begin{figure}[!htb]
\begin{center}
\includegraphics[height=4.5cm]{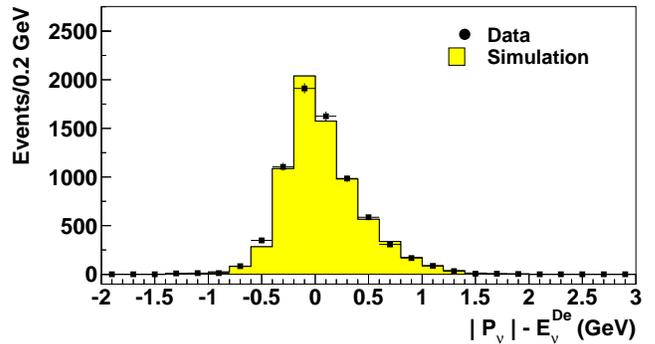}\\
\caption{The difference between the two neutrino energy estimates
described in the text for continuum-subtracted data 
and simulated $\BB$ events for the \Denu\ control sample.}
\label{fig:nures}
\end{center}
\end{figure}

The quality of the neutrino reconstruction is evaluated using a
control sample (\Denu) consisting of the decays
$\Bbar\to \Dz\electron\nub(X)$, where kinematic criteria result
in the $X$ system typically being no more than a 
$\pi$ or $\gamma$ from a $\Dstar\to\Dz X$ transition.  
The $\Dz$ is reconstructed in
the $\Km\pip$ decay mode and we require $|{\bf p}_{\Dz}|>0.5\gev$
and $E_e>1.4\gev$.  The $\Dz\,e$ combination must satisfy
$-2.5<\cos\theta_{B\cdot De}<1.1$, where $\cos\theta_{B\cdot De}
=(2E_B E_{De}-m_B^2-m_{De}^2)/(2|{\bf p}_B| |{\bf p}_{De}|)$ is
the cosine of the angle between the vector momenta of the $\Bbar$ and
the $\Dz\,e$ system assuming the only missing
particle in the $\Bbar$ decay was a single neutrino.  
After the combinatorial background is subtracted using
$\Dz$ mass sidebands, the selected sample 
consists primarily ($\simeq 95\%$) of
$\Bbar\to \Dz e\nub$ and $\Bbar\to \Dstar e\nub$ decays.
The control sample selection makes no requirements on the
other \B\ in the event, and can therefore be used
to study the impact of the modeling of the other \B\ on the neutrino
reconstruction.  Since the unreconstructed $X$ system
in the $\Bbar\to \Dz\electron\nub(X)$ decays carries away little energy,
a good estimate (r.m.s.$\sim 0.2\gev$) of the neutrino energy 
can be obtained from the known $\Bbar$ energy and the measured
$\Dz$ and $e$ energies, $E_{\nu}^{\rm De}$.  
A second estimate of the neutrino energy
is constructed from the visible momentum as described previously.  
Subtracting the
first estimate from the second gives the distribution shown in
Fig.~\ref{fig:nures}, where the criteria (1)--(5) described above
have been imposed.  We find good agreement between data and MC;
the average (r.m.s.) is
$0.066\gev$ ($0.366\gev$) for data and $0.072\gev$ ($0.365\gev$) for
simulated events.

\indent The \Denu\ control sample is also used to improve the modeling of
the $\btocenu$ decays.
After relaxing the $\cos\theta_{B\cdot De}$ requirements and
subtracting continuum and combinatorial backgrounds, 
we perform a binned $\chi^2$ fit to the $\Denu$ sample
in the variables $|{\bf p}_{D}|$, $E_e$ and $\cos\theta_{B\cdot De}$.
The fit determines scale factors for the MC
components $\Bbar\to De\nub$, $\Bbar\to \Dstar e\nub$ and other
contributions (85\%\ of which are decays to $D^{**}$ states),
while keeping the total $\btocenu$ branching fraction 
fixed to the measured value~\cite{ref:Elmoments}.
The fit increases the $\Bbar \to De\nub$ and $\Bbar \to \Dstar e\nub$ 
branching fractions to 2.29\% and 6.02\% (2.48\% and 6.52\%)
for neutral (charged) $\B$ mesons, respectively, 
while decreasing the remaining contributions.  
By design, these revised branching fractions 
respect isospin symmetry and are used in the determination of the background.

Two control samples are used
to reduce the sensitivity of the efficiency and background estimates
to details of the simulation: the \Denu\ control sample described above,
but with $E_e>2.0\gev$;
and events satisfying the normal selection criteria but having
$\shmax>4.25\GEV$, a sample with $<5\%$ signal decays.  
Efficiencies $\epsilon_{\Denu}^{\mathrm{data}}$ and $\epsilon_{\Denu}^{\mathrm{MC}}$
are calculated separately in
data and MC as the ratio of \Denu\ candidates satisfying criteria
(1)--(5) to the total \Denu\ sample.  
The $\btouenu$ signal efficiency is multiplied by
the ratio of these efficiencies 
to reduce sensitivity to details of the simulation.
The $\shmax>4.25\GEV$ sideband region is used to normalize the simulated $\shmax$
distribution to the data, reducing sensitivity to background
normalization uncertainties.  
 
We determine a partial branching fraction 
$\Delta\mathcal{B}(\tilde{E},\shmaxz)=\mathcal{B}(\btouenu)f_u$, 
unfolded for detector effects.
The acceptance,
$f_u$, is the fraction of \btouenu\ decays in the region of interest,
$\tilde{E}_e>2.0\gev$ and $\shmaxz<3.5\GEV$, where
$\tilde{E}_e$ and $\shmaxz$ are the true (generated) values in the
\B\ meson rest frame.  Slightly lower values are accepted for 
$\tilde{E}_e$ than for $E_e$ to account for the boost 
of the $B$ meson and to increase $f_u$.
The efficiency times acceptance for $\btouenu$ decays can be written as
$\epsilon_u=\epsilon_{\,\mathrm{sig}}f_u+\epsilon_{\,\overline{\mathrm{sig}}}(1-f_u)$,
where $\epsilon_{\,\mathrm{sig}}\ (\epsilon_{\,\overline{\mathrm{sig}}})$
is the efficiency for an event inside (outside) the 
region of interest to be reconstructed and pass our selection criteria.
We calculate the partial branching fraction as follows:
\begin{equation}
\label{eq:upbf}
\Delta\mathcal{B}
= \frac{N_{\mathrm{cand}}\ -\ M_{\mathrm{bkg}}
 \ \frac{N_{\mathrm{side}}}{M_{\mathrm{side}}}}{
   2\ N_{\BB}\ \frac{\epsilon_{\Denu}^{\mathrm{data}}}{\epsilon_{\Denu}^{\mathrm{MC}}}\ \epsilon_{\,\mathrm{sig}}}
    \ \left[ 1 + \frac{1-f_u}{f_u} \frac{\epsilon_{\,\overline{\mathrm{sig}}}}{\epsilon_{\,\mathrm{sig}}}\right]^{-1},
\end{equation}
where $N_{\mathrm{cand}}$ and $N_{\mathrm{side}}$ refer to the number of
candidates in the signal and $\shmax$ sideband regions of the
data, $M_{\mathrm{bkg}}$ and $M_{\mathrm{side}}$ refer to background
in the signal region and the yield in the sideband region in simulated
events and $2 N_{\BB}$ is the number of $B$ mesons produced from
$\FourS\to\BB$ decays.  
Since the resulting ratio of
$\epsilon_{\,\overline{\mathrm{sig}}}/{\epsilon_{\,\mathrm{sig}}}$
is small, 
$\Delta\mathcal{B}$ depends only weakly on the model used to
determine $f_u$.

\begin{figure}[!htb]
\begin{center}
\includegraphics[height=9cm]{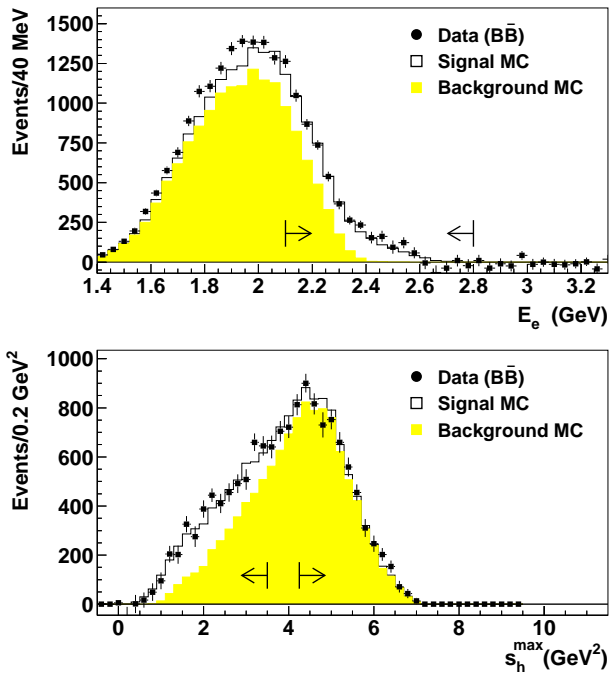}\\
\caption{The electron energy, $E_e$, 
and $\shmax$ spectra in the $\FourS$ frame
for continuum-subtracted data and simulated \BB\ events satisfying all
selection requirements except for the variable shown.
The arrows denote the signal (and sideband) region in $E_e$ and $\shmax$.
Note that $\tilde{E}_e \neq E_e$ (see text).}
\label{fig:plepstar}
\end{center}
\end{figure}

\indent Fig.~\ref{fig:plepstar} shows the electron energy and $\shmax$
distributions after cuts have been applied to all variables except the
one being displayed.  The discrepancy observed between data
and MC for $E_e<1.95\gev$ is covered by the systematic
error on the $\btocenu$ modeling.
The yields and efficiencies are given in
Table~\ref{tab:yields}.  We find
\begin{equation}
\Bunfb
\end{equation}
where the uncertainties are statistical and systematic, respectively.
Alternative values of $\Delta \mathcal{B}$ are obtained using different 
electron energy requirements: $\Bunfa$ and $\Bunfc$.
\begin{table}[htb]
\caption{Yields and efficiencies from data and simulation.  
All uncertainties are statistical except for $N_{\BB}$ where
systematics are included.  Efficiencies are quoted in units of $10^{-4}$.}
\begin{center}
\begin{tabular}{|c|c|c|c|c|c|} \hline
 $N_{\mathrm{cand}}$ & $N_{\mathrm{side}}$ &
 $\epsilon_{\Denu}^{\mathrm{data}}$ &
 \multicolumn{3}{c|}{$N_{\BB}\ (10^{6})$} \\
 $5130\pm 150$ & $6152\pm 130$ & $902\pm 39$
 & \multicolumn{3}{c|}{$88.35\pm 0.97$} \\ \hline
 $M_{\mathrm{bkg}}$  & $M_{\mathrm{side}}$
 & $\epsilon_{\Denu}^{\mathrm{MC}}$ & $\epsilon_{\,\mathrm{sig}}$
 & $\epsilon_{\,\overline{\mathrm{sig}}}$ & $f_u$ \\
 $3176\pm 35$ & $6423\pm 49$ & $906\pm 19$ & $256\pm 3$
 & $2.7\pm 0.2$ & $0.174$ \\ \hline
\end{tabular}
\end{center}
\label{tab:yields}
\end{table}

\indent Systematic uncertainties are assigned for the modeling of the signal
$\btouenu$ decays, background and
detector response.  The leading sources of uncertainty are listed in
Table~\ref{tab:syserr}.  Uncertainties from the simulation of
charged particle tracking, neutral reconstruction, charged 
particle identification, and the energy deposition by $\KL$
were evaluated from studies comparing data and simulation.
Radiation in the decay process was simulated using PHOTOS~\cite{ref:photos};
comparisons with the analytical result of Ref.~\cite{ref:ginsberg} were
used to assess the systematic uncertainty.  The uncertainty due to
bremsstrahlung in the detector was evaluated using the method of
Ref.~\cite{ref:Elmoments}.  
The uncertainty in modeling the background was
first evaluated by varying the total $\btocenu$,
$\Bbar\ra De\overline{\nu}$ and $\Bbar\ra \Dstar e\overline{\nu}$ rates
within their measured range.
Furthermore, the form factors for $\Bbar\ra De\overline{\nu}$~\cite{ref:dff} 
and $\Bbar\ra \Dstar e\overline{\nu}$~\cite{ref:dsff} were varied
within their uncertainties, and the composition of the $D^{**}$ states
was modified to include only narrow resonances, broad resonances, or
Goity-Roberts decays~\cite{ref:Goity}; the effect
of these variations is reduced by the fit to the \Denu\ control sample.
The modeling of $D$ decays was varied based on the measurements reported in
Ref.~\cite{ref:pdg2004};  the variation in the $D\to\KL X$  
branching fractions dominates the uncertainty.
The modeling of \btouenu\ decays is sensitive to the resonance structure at
low mass.  The branching fractions of 
$\Bbar\to (\pi,\,\rho,\,\omega,\,\eta,\,\eta^{\prime})e\nub$ were varied
as follows: $\pi$: $\pm 30\%$; $\rho$: $\pm 30\%$; $\omega$: $\pm 40\%$;
simultaneously $\eta$ and $\eta^{\prime}$: $\pm 100\%$.
\begin{table}[htb]
\caption{Uncertainties on $\Vub$ and $\Delta\mathcal{B}$.}
\begin{center}
\begin{tabular}{|l|c|c|} \hline
Source & $\sigma(\Vub)/\Vub\,(\%)$ & $\sigma(\Delta\mathcal{B})/\Delta{\mathcal{B}}\,(\%)$ \\ 
\hline \hline
Tracking             &  $\pm 0.8$ & $\pm 1.5$ \\
Neutrals             &  $\pm 1.7$ & $\pm 3.4$ \\
Electron ID          &  $\pm 0.5$ & $\pm 1.0$ \\
Hadron ID            &  $\pm 1.0$ & $\pm 2.0$ \\
Bremsstrahlung       &  $\pm 1.0$ & $\pm 2.0$ \\
$\KL$                &  $\pm 1.3$ & $\pm 2.6$ \\
$N_{B\overline{B}}$  &  $\pm 0.6$ & $\pm 1.1$ \\
Radiation            &  $\pm 1.9$ & $\pm 3.8$ \\
$\btocenu$ modeling  &  $\pm 2.5$ & $\pm 5.0$ \\ 
$\btouenu$ resonances&  $\pm 2.2$ & $\pm 4.4$ \\ 
Statistical          &  $\pm 4.7$ & $\pm 9.3$ \\
\hline
Total experimental   & $\pm 6.7$  & $\pm 13.3$ \\ 
\hline
Heavy quark parameters & ${}^{+14.6}_{-10.6} $ & $\pm 1.5$ \\ 
Theoretical          & $\pm 6.3 $ &  \\ 
\hline
\end{tabular}
\end{center}
\label{tab:syserr}
\end{table}

\indent We extract
$\Vub=[\Delta\mathcal{B}/(\Delta\zeta\ \tau_B)]^{1/2}$
using $\tau_B=1.604\pm 0.023\ps$~\cite{ref:pdg2004}.
The normalized partial rate,
$\Delta \zeta$, computed in units of $\Delta\Gamma/\Vub^2$,
is taken from Ref.~\cite{ref:Notebook},
in which 
the leading terms in the HQE of the 
$\btouenu$ spectra are computed at next-to-leading order, 
and power corrections are included
at $\mathcal{O}(\alpha_S)$ for the leading shape function (SF) and at 
tree level for subleading SFs.
The values used for the heavy quark parameters,
$m_b=4.61\pm 0.08\gev$ and $\mu_{\pi}^2=0.15\pm 0.07\GEV$, with a
correlation coefficient of $-0.4$,
are based on fits to $\btoclnu$ moments~\cite{ref:OPEfits}, 
translated to the shape-function scheme of Ref.~\cite{ref:Neubert-new-4}.  

We find $\Vubb$ for $\tilde{E}_e>2.0\gev$, where the errors represent
experimental, heavy quark parameters,
and theoretical uncertainties, respectively.
The latter include estimates of the effects of
subleading SFs~\cite{ref:subsf}, 
variations in the matching scales used in the calculation,
and weak annihilation~\cite{ref:WA}.  No uncertainty is assigned for
possible quark-hadron duality violation.
The determination of \Vub\ is limited primarily by our knowledge of 
$m_b$.  An approximate dependence is
$|V_{ub}(m_b)|=|V_{ub}(m_0)|\left(1+7\,(m_b-m_0)/m_0\right)$, where
$m_0=4.61\gev$.
The sensitivity to higher moments of the SF is weak; the change in 
\Vub\ when varying $\mu_{\pi}^2$ from 0.03 to 0.35\GEV\ with $m_b$ fixed 
is $2\%$, and the impact of using alternative SF 
parameterizations~\cite{ref:bsgfits} is $< 2\%$.
The overall precision on the above result 
surpasses that of Refs.~\cite{ref:endp-meas} and \cite{ref:mx-meas}, 
but is comparable to
determinations of \Vub\ which have become available while this
paper was nearing completion~\cite{ref:HFAG}.

\section{Acknowledgments}
\label{sec:Acknowledgments}
We are grateful for the excellent luminosity and machine conditions
provided by our \pep2\ colleagues, 
and for the substantial dedicated effort from
the computing organizations that support \babar.
The collaborating institutions wish to thank 
SLAC for its support and kind hospitality. 
This work is supported by
DOE
and NSF (USA),
NSERC (Canada),
IHEP (China),
CEA and
CNRS-IN2P3
(France),
BMBF and DFG
(Germany),
INFN (Italy),
FOM (The Netherlands),
NFR (Norway),
MIST (Russia), and
PPARC (United Kingdom). 
Individuals have received support from CONACyT (Mexico), 
Marie Curie EIF (European Union),
the A.~P.~Sloan Foundation, 
the Research Corporation,
and the Alexander von Humboldt Foundation.

Finally, we would like to thank the many theorists with 
whom we have had valuable
discussions, and further thank M. Neubert, B. Lange and G. Paz for making
available for our use a computer code implementing their calculations.

\end{document}